\documentclass[prb,twocolumn]{revtex4}

\usepackage{epsf}
\usepackage{epsfig}
\usepackage{textcomp}
\usepackage{dcolumn}
\usepackage{bm}

\begin{document}


\title{Exact-Exchange Spin-Current Density-Functional Theory}

\author{Stefan Rohra and Andreas G\"orling}
\affiliation{Lehrstuhl f\"ur Theoretische Chemie,
Universit\"at Erlangen-N\"urnberg, Egerlandstr. 3, D-91058 Erlangen, Germany}

\date{\today}

\begin{abstract}
A spin-current density-functional theory (SCDFT) 
is introduced, which takes into account currents of the 
spin-density and thus currents 
of the magnetization in addition 
to the electron density, the non-collinear spin-density, 
and the density current,
which are considered in standard current spin-density-functional theory. 
An exact exchange Kohn-Sham formalism based on SCDFT is 
presented, which represents a general framework for the treatment of 
magnetic and spin properties. 
As illustration an oxygen atom in a magnetic field is treated 
with the new approach.

\end{abstract}


\maketitle

\def\angst{\,\text{\AA}}

\newcommand{\elmat}[3]{\langle {#1} | {#2} |{#3} \rangle}

\def\RR{({\bf r})}
\def\RB{{\bf r}}       
\def\RP{({\bf r}')}
\def\RPP{({\bf r}'')}
\def\phiG#1{\phi^{\Gamma\!,\gamma}_{#1}}
\def\phiGT#1{{\phi^{\Gamma\!,\gamma}_{#1}}^T\!\!}
\def\phiU#1{\phi^{\Upsilon\!,\upsilon}_{#1}}
\def\phiGP#1{\phi^{\Gamma'\!,\gamma'}_{#1}}
\def\phiGPT#1{{\phi^{\Gamma'\!,\gamma'}_{#1}}^T\!\!}
\def\phiGM#1{\phi^{\Gamma_M,\gamma}_{#1}}
\def\phiGMT#1{{\phi^{\Gamma_M,\gamma}_{#1}}^T\!\!}
\def\phiGm#1{\phi^{\Gamma\!,\mu}_{#1}}
\def\phiGmT#1{{\phi^{\Gamma\!,\mu}_{#1}}^T\!\!}
\def\nG#1{n^{\Gamma}_{#1}}
\def\nGP#1{n^{\Gamma'}_{#1}}
\def\nGM#1{n^{\Gamma_M}_{#1}}
\def\phiL#1{\phi^{\Lambda,\lambda}_{#1}}
\def\phiLT#1{{\phi^{\Lambda,\lambda}_{#1}}^T\!\!}
\def\nL#1{n^{\Lambda}_{#1}}
\def\gG{g^{\Gamma}}
\def\gGP{g^{\Gamma'}}
\def\gGM{g^{\Gamma_M}}
\def\gL{g^{\Lambda}}
\def\eG#1{\varepsilon^{\Gamma}_{#1}}

\def\coulP{|{\bf r}'-{\bf r}''|}
%

%
%

Density-functional methods \cite{yang89}, 
in particular Kohn-Sham (KS) methods, 
have developed into the most widely used approach
to investigate electronic structures of molecules, clusters, surfaces, 
or solids. However, magnetic properties and generally properties and
quantities related to currents cannot be described with the available
KS methods relying on standard spin-density-functional theory, 
which is based on
the spin-density alone. 
In order to treat magnetic effects due to both spin and currents
within a density-functional theory (DFT) framework, 
Vignale and Rasolt therefore introduced a current 
spin-density-functional formalism almost 
20 years ago \cite{vignale87,vignale88}. Despite the fact
that magnetic effects are ubiquitous and of paramount importance, see, e.g.,
magnetooptics, spintronics, or nuclear magnetic resonance spectroscopy, 
current spin-density-functional theory (CSDFT) did not play a significant 
role in practice so far. The  reason is that
sufficiently accurate approximations for the required but not
exactly known current spin-density functionals for 
exchange and correlation are not available. 

In recent years exact-exchange (EXX) KS methods 
have been developed, which treat 
exchange energy and potential exactly and therefore do not require any 
approximate exchange functionals \cite{gorling05}. 
These EXX KS methods yield band structures 
for semiconductors and generally one-particle spectra of electronic systems,
which are strongly improved compared to methods based on the LDA or 
generalized gradient approximations 
\cite{gorling05, stadele97}. A development of EXX 
current density-functional methods therefore seems highly promising 
\cite{kurth05}. 

CSDFT depends on seven variables: the electron density, 
the x-, y-, z-components of the spin-density, and the 
x-, y-, z-components of the paramagnetic density current
\cite{vignale87,vignale88}. 
However, as demonstrated below, in general, 
also currents of the spin-density,
i.e., currents of the magnetization, may occur 
in the presence of magnetic fields. For a complete 
description of magnetic effects it therefore 
seems desirable to go beyond CSDFT and to include also 
the x-, y-, z-components
of the currents of the x-, y-, z-components of the spin-density, i.e., 
to include 9 more variables leading to a total of 16 variables. The
resulting formalism shall be called spin-current density-functional theory 
(SCDFT) as opposed to current spin-density-functional theory in order to
emphasize that it depends on spin-current densities (The term spin-current
density from now on shall include besides actual spin-current densities also
standard current densities 
as well as non-collinear 
spin-densities including the standard electron density). 
In this Letter we firstly introduce
the basic SCDFT, then present an exact exchange KS formalism based on SCDFT, 
and finally consider as an illustrative example an oxygen atom
in a magnetic field.

The Hamiltonian operator of an electronic system in a magnetic field is
given by
\begin{eqnarray}
\hat{H} \!\! 
&=&\hat{T} + \hat{V}_{ee} + \sum_{i=1}^N \bigg [ v_{ext}({\bf r}_i) 
+ {\textstyle \frac{1}{2}} {\bf p}_i \!\cdot\! {\bf A}({\bf r}_i)
+ {\textstyle \frac{1}{2}} {\bf A}({\bf r}_i) \!\cdot\! {\bf p}_i
\nonumber \\[-0.5em]
&& \;\;\;\;\;\;\;\;\;\;\;\;\;\;\;\;\;\;\;\;\;\;\;\;\;\;\;\;\;\;\;\;\;\;\;
+ {\textstyle \frac{1}{2}} 
{\bf A}({\bf r}_i) \!\cdot\! {\bf A}({\bf r}_i)
+  {\textstyle\frac{1}{2}} \, \boldsymbol{\sigma} \!\cdot\! {\bf B}({\bf r}_i) 
\bigg ]
\nonumber \\
&=&\hat{T} + \hat{V}_{ee} + \int \! d{\bf r} \;\; \boldsymbol{\Sigma}^T \;
{\bf V}({\bf r}) \; \hat{\bf J}({\bf r}) \,.
\label{ham_all_el}
\end{eqnarray}
In Eq. [\ref{ham_all_el}], $\hat{T}$ and $\hat{V}_{ee}$ are the operators of
the kinetic energy and the electron-electron repulsion, ${v}_{ext}$ is
the external electrostatic potential, usually the potential of the nuclei, 
${\bf B}$ designates a magnetic field
with accompanying vector potential ${\bf A}$. By ${\bf r}_i$ the position of
the $i$-th electron is denoted, by ${\bf p}_i$ the corresponding canonical
momentum operator. The vector $\boldsymbol{\sigma}$ contains the
Pauli spin matrices. The sum in the first line of Eq. [\ref{ham_all_el}] runs
over all $N$ electrons. The vector $\boldsymbol{\Sigma}$ has four components,
${\Sigma}_0$ being a 2x2 unit matrix and ${\Sigma}_1$, ${\Sigma}_2$, and
${\Sigma}_3$ being the Pauli spin matrices ${\boldsymbol\sigma}_x$, 
${\boldsymbol\sigma}_y$, and ${\boldsymbol\sigma}_z$. The four components 
of the vector $\hat{\bf J}({\bf r})$ are the density operator 
$\sum_{i=1}^N \; \delta({\bf r}-{\bf r}_i) = \hat{J}_{0}({\bf r})$
and the x-, y, z- components of the current operator   
$\left(\frac{1}{2}\right) 
\sum_{i=1}^N \;\; p_{x,i} \; \delta({\bf r}-{\bf r}_i) \;+\;
\delta({\bf r}-{\bf r}_i) \; p_{x,i} =  \hat{J}_{1}({\bf r})$ etc. 
($p_{x,i}=-i \partial/\partial x_i$ denotes the 
momentum operator of the $i$-th electron in 
x-direction.) The 4x4 matrix ${\bf V}({\bf r})$ is composed of matrix elements
${V}_{\mu\nu}({\bf r})$ with $\mu,\nu= 0,1,2,3$ and is given by
\begin{equation}
\mathbf{V}({\bf r})
=\!
\left( \!\!  \begin{array}{cccc}
v_{ext}({\bf r}) \!+\!  \frac{{\bf A}^2({\bf r})}{2} \!&\! 
A_x({\bf r}) \!&\! A_y({\bf r}) 
\!&\! A_z({\bf r}) \\
\frac{B_x({\bf r})}{2} \!\!\!&\!\!\! 0 
\!\!\!&\!\!\! 0 \!\!\!&\!\!\! 0 \\
\frac{B_y({\bf r})}{2} \!\!\!&\!\!\! 0
\!\!\!&\!\!\! 0 \!\!\!&\!\!\! 0 \\
\frac{B_z({\bf r})}{2} \!\!\!&\!\!\! 0
\!\!\!&\!\!\! 0 \!\!\!&\!\!\! 0
\end{array} \! \right) \,.
\label{V}
\end{equation}

The total energy $E$ of the system is given by
\begin{eqnarray}
E \! &=&  {T} + {V}_{ee} 
+ \sum_{\mu\nu} \int\!\! d{\bf r} \;
{V}_{\mu\nu}({\bf r}) \; \rho_{0,\mu\nu}({\bf r})
\nonumber \\
 &=& {T} + {V}_{ee} 
+  \int \!\! d{\bf r} \;\;
{\bf V}^T({\bf r}) \; \boldsymbol{\rho}_0({\bf r}) \,. 
\label{energy}
\end{eqnarray}
In Eq. [\ref{energy}], ${T}$ and ${V}_{ee}$ denote the kinetic and the
electron-electron repulsion energy, respectively, and  $\boldsymbol{\rho}_0$
is the ground state spin-current density 
consisting of 16 components ${\rho}_{0,\mu\nu}$ 
with $\mu,\nu = 0,1,2,3$. The component ${\rho}_{0,00}$ is the regular
ground state electron density, ${\rho}_{0,\mu 0}$ with $\mu = 1,2,3$ represent
the x-, y-, z-components of the spin-density, ${\rho}_{0,0 \nu}$ with 
$\nu = 1,2,3$ represent the x-, y-, z-components of the paramagnetic
current of the electron density, while  ${\rho}_{0,\mu\nu}$ 
with $\mu,\nu = 1,2,3$
represent the components of the paramagnetic currents of the
spin-density, i.e., currents of the magnetization. The corresponding physical
currents consisting of the contributions 
$\rho_{0,\mu\nu}({\bf r})+ \rho_{0,\mu0}({\bf r}) A_{\nu}({\bf r})$  
describe, via the continuity equation, changes of the magnetization in time 
and thus are meaningful physical quantities.
The spin-current density $\boldsymbol{\rho}_0$ can be 
considered as a 4x4 matrix or, like in the second line of Eq. [\ref{energy}], 
as a sixteen-component vector with superindex $\mu\nu$. Similarly, 
${\bf V}({\bf r})$ can be treated as 4x4 matrix, like in the second line of 
Eq. [\ref{ham_all_el}], or as sixteen-component vector, 
like in the second line of Eq. [\ref{energy}].

The reason to introduce the matrix ${\bf V}({\bf r})$ and to rewrite the
Hamiltonian operator according to the second line of Eq. [\ref{ham_all_el}] is
that this leads to the expression [\ref{energy}] for the total
energy $E$. Starting from this expression for $E$ it is straightforward to
generalize the definitions and derivations of standard
DFT. 
A Hohenberg-Kohn functional $F[\boldsymbol{\rho}]$ can be defined via 
a generalized constrained-search \cite{higuchi04,nagy05} by
\begin{equation}
F[\boldsymbol{\rho}] \; = \;
\raisebox{-1.6ex}{$\stackrel{\displaystyle{\text{Min}}}
{\scriptstyle{\Psi \rightarrow \boldsymbol{\rho}({\bf r}) }}$}
\langle \Psi | \hat{T} \;+\; \hat{V}_{ee} |
\Psi \rangle \,.
\label{constrained_search}
\end{equation}
The minimization [\ref{constrained_search}] runs over all wave functions 
yielding the spin-current density $\boldsymbol{\rho}$, i.e., the wave
functions are constrained to not only yield a given reference electron
density, like in standard DFT, but to yield the 16 components $\rho_{\mu\nu}$
of the spin-current density. The minimizing wave function 
$\Psi[\boldsymbol{\rho}]$ of (\ref{constrained_search}) is a functional 
of the spin-current density $\boldsymbol{\rho}$. The
minimizing wave function $\Psi[\boldsymbol{\rho}_0]$ for the ground state 
spin-current density $\boldsymbol{\rho}_0$ is the ground state wave function
$\Psi_0$ of the system, $\Psi_0 = \Psi[\boldsymbol{\rho}_0]$, 
i.e., in case of the presence of magnetic fields,
the ground state current-spin density determines via the constrained-search 
(\ref{constrained_search})
the ground state wave function and thus all ground state properties of the 
system. The latter statement represents a Hohenberg-Kohn theorem for SCDFT.

A Kohn-Sham wave function $\Phi[\boldsymbol{\rho}]$ corresponding to a
wave function $\Psi[\boldsymbol{\rho}]$ is defined as the minimizing wave 
function in the constrained search minimization 
\begin{equation}
T_s[\boldsymbol{\rho}] \; = \;
\raisebox{-1.6ex}{$\stackrel{\displaystyle{\text{Min}}}
{\scriptstyle{\Phi \rightarrow \boldsymbol{\rho}({\bf r}) }}$}
\langle \Phi | \hat{T} |
\Phi \rangle
\label{constrained_search_Ts}
\end{equation}
yielding the functional $T_s[\boldsymbol{\rho}]$
of the noninteracting kinetic energy. As usually, exchange and correlation
energy 
are defined as $E_x[\boldsymbol{\rho}] =  \langle \Phi[\boldsymbol{\rho}] 
| \hat{V}_{ee} | \Phi[\boldsymbol{\rho}] \rangle - U[\rho_{00}]$ 
and $E_c[\boldsymbol{\rho}] =  \langle \Psi[\boldsymbol{\rho}] 
| \hat{T} \;+\; \hat{V}_{ee} | \Psi[\boldsymbol{\rho}] \rangle 
- \langle \Phi[\boldsymbol{\rho}] 
| \hat{T} \;+\; \hat{V}_{ee} | \Phi[\boldsymbol{\rho}] \rangle$ with the
Coulomb energy $U[\rho_{00}]$ given by the standard expression 
$\frac{1}{2} \int \!\! d{\bf r} d{\bf r}' 
\frac{\rho_{00}({\bf r}) \rho_{00}({\bf r}')}{|{\bf r} -{\bf r}'|}$ depending
only on $\rho_{00}$.

The many-electron KS equation corresponding to a Schr\"odinger equation for
real electrons 
with Hamiltonian operator [\ref{ham_all_el}] 
is determined by the KS Hamiltonian operator
\begin{eqnarray}
\hat{H}_s \!\! &=& \hat{T} 
+ \int \! d{\bf r} \;\; \boldsymbol{\Sigma}^T \;
{\bf V}_s({\bf r}) \; \hat{\bf J}({\bf r})
\label{KS-Hamiltonian}
\end{eqnarray}
with the 4x4 matrix ${\bf V}_s$ representing the KS potential,
given by
${\bf V}_s({\bf r}) = {\bf V}({\bf r}) + {\bf U}({\bf r})
+ {\bf V}_{xc}({\bf r})$. The Coulomb potential ${\bf U}$
and the exchange-correlation potential ${\bf V}_{xc}$ can be represented by
4x4 matrices; within the matrix ${\bf U}$ only the component 
$U_{00}({\bf r}) = 
\int \!\! d{\bf r}' \frac{\rho_{00}({\bf r}')}{|{\bf r} -{\bf r}'|}$
is different from zero and is the standard Coulomb potential. 
By straightforward generalization of the arguments
of the standard KS formalism \cite{yang89}, i.e., by invoking the variational
principle for the KS system and the real electron system and by comparing the
resulting Euler equations \cite{higuchi04,nagy05},
it can be deduced that the components 
${\bf V}_{xc,\mu\nu}$ of the 
exchange-correlation potential ${\bf V}_{xc}$ are given by
\begin{eqnarray}
V_{xc,\mu\nu}({\bf r}) = 
\frac{\delta E_{xc}}{\delta \rho_{\mu\nu}({\bf r})} \,.
\label{Vxc}
\end{eqnarray}
In the functional derivatives (\ref{Vxc}) all 16 components of $\rho_{\mu\nu}$
are treated as independent. As a result all 16 components
$V_{xc,\mu\nu}$ and subsequently $V_{s,\mu\nu}$, including components (or
combinations thereof) corresponding to gauge transformations, are determined
within the presented formalism. The gauge of all quantities is determined by
the gauge of the external vector potential ${\bf A}$. (A modified version of
the formalism chooses the Coulomb gauge throughout and treats only the
transversal components of the spin-currents \cite{rohra06b}.)
While the KS 
wave function $\Phi[\boldsymbol{\rho}_0]$, corresponding to the ground
state $\Psi_0 = \Psi[\boldsymbol{\rho}_0]$ of a real system, is
uniquely defined by the constrained-search (\ref{constrained_search_Ts}),
the functional derivatives (\ref{Vxc}) 
uniquely define the exchange-correlation potential ${\bf V}_{xc}$ and
subsequently yield the effective KS potential
${\bf V}_s$ 
if they are evaluated at $\boldsymbol{\rho}_0$. Like in the standard 
KS formalism, 
it is assumed that the functional derivatives (\ref{Vxc}) 
exist and that
the spin-current density $\boldsymbol{\rho}_0$ is 
noninteracting $v$-representable.

The components ${V}_{xc,\mu\nu}$ and subsequently 
${V}_{s,\mu\nu}$ with $\mu,\nu=1,2,3$, i.e., the components coupling 
to actual 
spin-currents, in general, differ from zero despite the fact 
that the corresponding
components in the external potential ${\bf V}$ of Eq. (\ref{V})
are zero. 
A similar situation occurs in the standard spin-density
formalism \cite{yang89}. 
If the spin-density of the system is spin-polarized then the KS 
exchange-correlation potential and the effective KS potential are
spin-polarized although the external potential is not spin-polarized.
Because the external potential does not contain a contribution coupling to
the spin-polarization it is, in principle, possible to treat spin-polarized
systems within the basic (non-spin-polarized) KS formalism. 
However, for approximate exchange-correlation functionals and thus in all 
practical applications, 
a spin-polarized treatment is more preferable for spin-polarized systems.
Similarly it is possible in principle, to treat
within the original current spin-density formalism  systems, 
which are subject to magnetic fields
because the external potential
does not couple to spin-currents, i.e.,
because ${V}_{\mu\nu}=0$ for $\mu,\nu=1,2,3$. 
However, in the presence of magnetic fields, in general, spin-currents occur
(see below) and therefore a treatment within the spin-current density
formalism presented here seems to be preferable.

If spin-orbit interactions are considered then the components 
${V}_{\mu\nu}$ for $\mu,\nu=1,2,3$ of the external potential no longer
are zero and the original current spin-density formalism is no longer
applicable and it is required to switch to the spin-current density
formalism presented here. Indeed a generalization of the
presented formalism that enables the treatment of spin-orbit
interactions is possible and will be published elsewhere \cite{rohra06}. 


In order to apply the presented 
SCDFT,
suitable 
exchange-correlation functionals are required. To that end, we concentrate 
on the exchange functionals and treat those exactly. 
Like in the standard KS formalism,
the evaluation of the exchange energy
is well-known from Hartree-Fock methods and 
straightforward. 
By generalizing the treatment of the standard exact-exchange
KS approach \cite{gorling05} 
also the exact exchange potential ${\bf V}_{x}$ of SCDFT 
becomes accessible. The chain rule of functional differentiation
yields
\begin{eqnarray}
&&\!\!\!\!\!\! v_{x,\mu\nu}({\bf r})  =
\nonumber \\ 
&&\!\!\!\!\!\! \sum_{\kappa\lambda} \!  \int\!\!d{\bf r}' 
\left [ \sum_{a}^{\text{occ.}} \!  \int\!\! d{\bf r}'' 
\; \frac{\delta E_{x}}{\delta \phi_a({\bf r}'')}  
\; \frac{\delta \phi_a({\bf r}'')}{\delta V_{s,\kappa\lambda}({\bf r}')} 
+ c.c. \right ] \!
 \frac{\delta V_{s,\kappa\lambda}({\bf r}')}{\delta \rho_{\mu\nu}({\bf r})} 
\,.
\nonumber \\
\label{EXX}
\end{eqnarray}
In Eq. (\ref{EXX}), $\phi_a$ stands for the KS orbitals, complex-valued
two-dimensional spinors, building the KS determinant. 
The right hand side (r.h.s) 
of Eq. (\ref{EXX}) due to the sum over $\kappa\lambda$
contains
16 terms. Furthermore, for each of the 16 components of ${\bf V}_{x}$ an
equation (\ref{EXX}) arises. These can be combined in the 16x16 matrix 
equation 
%
${\bf V}_{x}({\bf r})  \; = \;
\int\! d{\bf r}'  \; {\bf X}_s^{-1}({\bf r},{\bf r}') 
\; {\bf t}({\bf r}')$,
%
which is turned into 
\begin{equation}
\int\!\!d{\bf r}'  \;\; {\bf X}_s({\bf r},{\bf r}') 
\;\; {\bf V}_{x}({\bf r}')  \; = \;
{\bf t}({\bf r}) \,.
\label{EXX3}
\end{equation}
%
Here, ${\bf X}_s$ is the spin-current response function of the 
KS system, a 16x16 matrix with matrix elements
\begin{eqnarray}
{X_{s,\mu\nu,\kappa\lambda}({\bf r},{\bf r}')} \!\!\!&=&\!\!\!
\sum_a^{occ.} \!  \sum_s^{unocc.} \!\! 
\frac{\langle \phi_a | {\sigma}_{\mu} \hat{J}_{\nu}({\bf r})
                        | \phi_s \rangle \! 
       \langle \phi_s| {\sigma}_{\kappa} \hat{J}_{\lambda}({\bf r}')
                        | \phi_a \rangle }
{\varepsilon_a -\varepsilon_s} 
\nonumber \\[-0.5em]
&& \;\;\;\;\;\;\;\;\;\;\;\;\;\;\;\; + c.c. 
\label{Xs}
\end{eqnarray}
The components ${t}_{\mu\nu}$ of the 16-dimensional vector ${\bf t}$
are given by the terms in the square brackets on the right hand side of  
Eq. (\ref{EXX}). The functional derivative 
$\frac{\delta E_{x}}{\delta \phi_a({\bf r}'')}$ contained in the 
${t}_{\mu\nu}$ can be evaluated
straightforwardly because the exchange energy is known in terms of the KS
orbitals, see, e.g., Ref.[\onlinecite{gorling05}]. 
Eq. (\ref{EXX3}) represents the EXX equation of 
SCDFT; it is an integral equation, which needs to be solved in each
iteration of the SCDFT KS self-consistency process in order to obtain the 
exchange potential ${\bf V}_{x}$.

We implemented the presented EXX-SCDFT approach within a plane-wave solid
state code. The quantities $X_{s,\mu\nu,\kappa\lambda}$,
${v}_{x,\mu\nu}$, and ${t}_{\mu\nu}$ are
expanded into an auxiliary plane wave basis set. 
The $X_{s,\mu\nu,\kappa\lambda}$ then turn into matrices of the dimension of 
the auxiliary basis set, while the ${v}_{x,\mu\nu}$ and ${t}_{\mu\nu}$  
turn into vectors of the same dimension. 
The complete response function then turns into a matrix
consisting of 16x16 submatrices $X_{s,\mu\nu,\kappa\lambda}$ and the
complete exchange potential and r.h.s of Eq. (\ref{EXX3}) become vectors 
consisting of 16 subvectors ${v}_{x,\mu\nu}$ and ${t}_{\mu\nu}$.
The integral equation (\ref{EXX3}) turns into a matrix equation and can
easily be solved.

As an illustration we consider an oxygen atom subject to a constant 
magnetic field directed
along the z-axis. To that end, we put the oxygen atom in the centers of
simple cubic 
supercells of 9, 10, and 11 a.u. length. 
Plane wave cutoffs
of 14, 20, and 24 Rydberg for the orbitals and of 4, 6, and 7.2
Rydberg for the response function, the exchange potential, and the r.h.s of
Eq. (\ref{EXX3}) are chosen. All results shown below
are converged with respect to
supercell size and plane wave cutoffs and are invariant with
respect to gauge transformations of the external vector potential ${\bf A}$, 
within the numerical accuracy determined by the plane wave cutoffs. 
The oxygen 1s electrons are taken
into account via separable EXX pseudopotentials \cite{hoeck98}.

In the absence of magnetic fields and 
spin-orbit interactions the electronic ground state of oxygen is a 
nine-fold degenerate $^3P$-state of valence
configuration $(2s^2,2p^4)$. 
The spin and spatial angular momenta can be coupled
to give  $^3P_0$, $^3P_1$, and $^3P_2$ LS-coupled states. In the presence 
of a 
constant magnetic field in z-direction, the $^3P_2$-state with total magnetic
quantum number $M_J$=--2 is the ground state. 
This state is a one-determinantal state with occupied  $p$-orbitals 
$p_{-1,-1/2}$, $p_{0,-1/2}$, $p_{+1,-1/2}$, and $p_{-1,+1/2}$. (First and
second subscripts, $m_{\ell}$ and $m_{s}$, 
denote magnetic quantum numbers for spatial and spin angular
momentum, respectively.) 
A KS treatment of this state even in the limit of 
zero magnetic field leads to a symmetry breaking KS wave function and a
lifting of the degeneracy of the $p$-orbitals because of
the partially filled $p$-shell accompanied by a nonspherical but
cylindrical electron density \cite{gorling05,becke02}.
Correspondingly, the symmetry of the KS Hamiltonian operator
is reduced from spherical
to cylindrical symmetry around the z-axis. In the presence of the
magnetic field the latter is the actual symmetry of the system. 

\begin{figure}[h]
\includegraphics*[width=6.0cm]{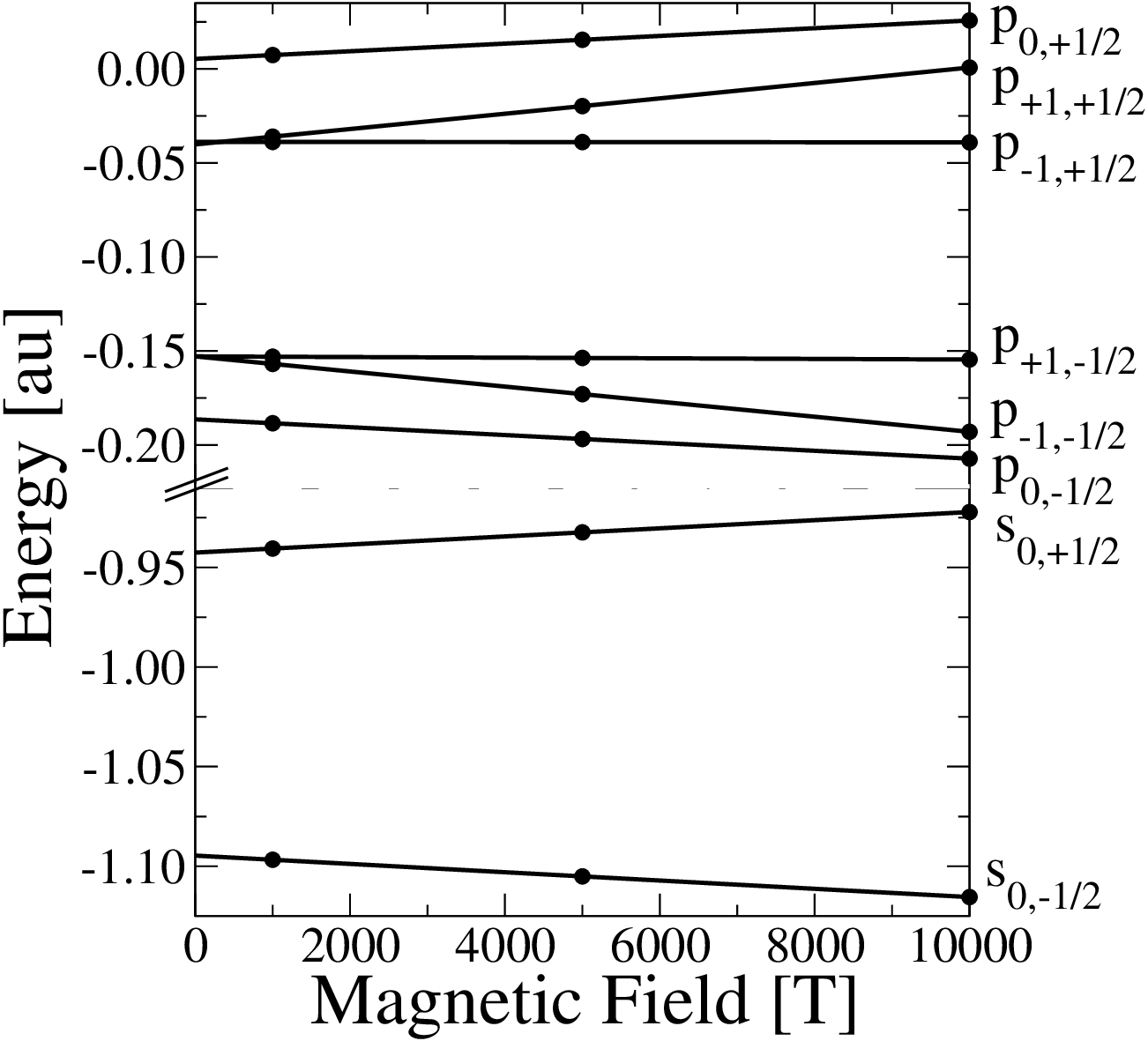}
\caption
{Eigenvalues of oxygen KS orbitals $s_{m_{\ell},m_s}$ and $p_{m_{\ell},m_s}$
as functions of magnetic field strength}
\label{eigenvalues}
\end{figure}

In the absence of spin-orbit interactions the change $\Delta E$ of orbital 
energies (and, below, of the total energy)  
due to the magnetic field of strength $B$
is described according to the Paschen-Back
formula $\Delta E \;=\; \beta \; B \; (m_{\ell} + 2 m_s)$  
not according to the Zeeman formula. Here $\beta$ stands for the 
Bohr magneton.
The values of $\Delta E/(\beta \; B)$ for the curves in 
Fig.\ \ref{eigenvalues} have to
equal and, indeed, equal $(m_{\ell} + 2 m_s)$. 
For the special case of the $^3P_2$-state with $M_J=-2$ the
Paschen-Back and the Zeeman formula $\Delta E \;=\; \beta \; g_J \; B \; M_J$
yield the same energy splitting with
respect to the magnetic field strength. It is thus possible to obtain 
from our calculation the  Land\'{e}  factor $g_J$ of oxygen despite 
the neglect of spin-orbit interactions. 
The resulting value of 1.5
agrees with the experimental one. This demonstrates that the presented SCDFT 
approach works correctly in practice in the sense that it correctly generates
spin and angular momenta and aligns them correctly with the magnetic field.
(Note that in our
plane wave supercell method spin and angular momenta as well as their coupling
and alignment are not enforced by the basis set and by occupation numbers 
but are an outcome of the SCDFT treatment.)  

\begin{figure}[h]
\includegraphics*[width=5.0cm]{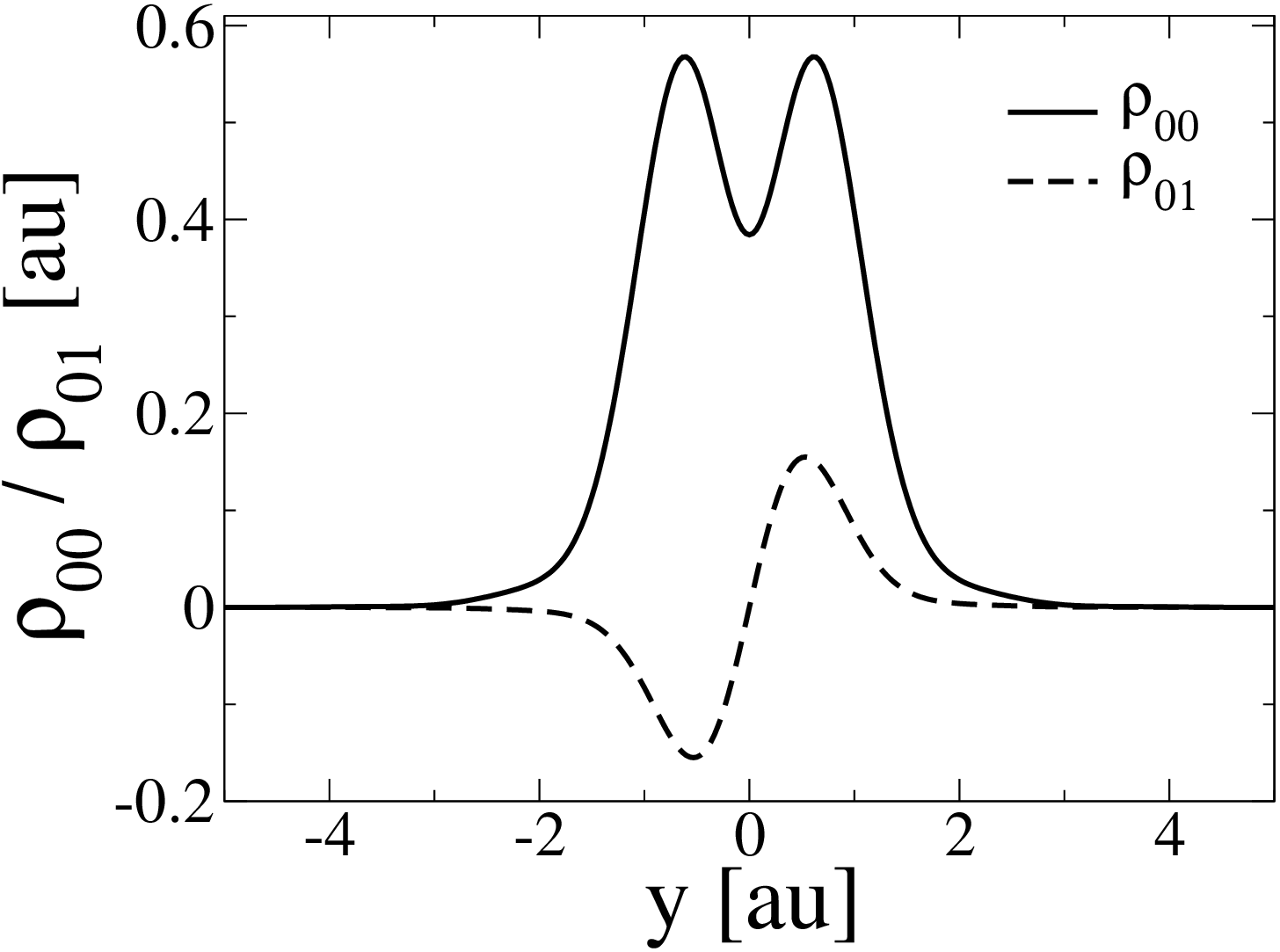}
\caption
{Electron density $\rho_{00}$ of oxygen and its current
$\rho_{01}$ in $x$-direction displayed along the 
$y$-axis for a magnetic field strength of 10000 T}
\label{rho00}
\end{figure}

\begin{figure}[h]
\includegraphics*[width=5.0cm]{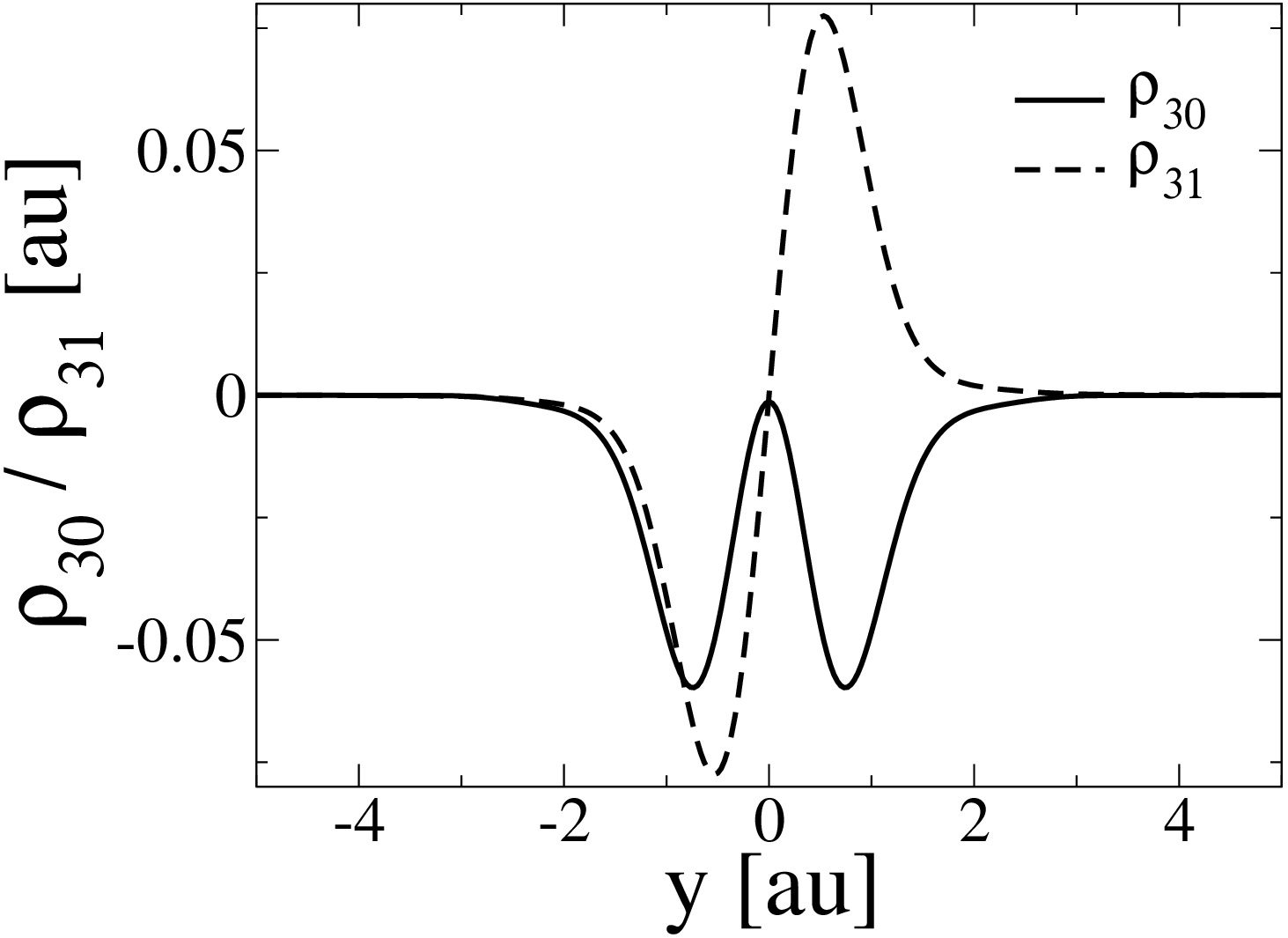}
\caption
{Spin density $\rho_{30}$ of oxygen and its current
$\rho_{31}$ in $x$-direction displayed along the 
$y$-axis for a magnetic field strength of 10000 T}
\label{rho30}
\end{figure}

In Fig.\ \ref{rho00} the electron density $\rho_{00}$ and its current
$\rho_{01}$ in $x$-direction are displayed along the $y$-axis.
As usual in pseudopotential calculations  $\rho_{00}$ exhibits a minimum at
the nucleus. The displayed $\rho_{01}$ corresponds to a current
around the $z$-axis. The currents $\rho_{02}$ and $\rho_{03}$ are zero. Thus 
the magnetic field correctly aligns the spatial angular momentum along the 
$z$-axis. In Fig.\ \ref{rho30} 
the spin-density $\rho_{30}$, i.e., the magnetization
along the $z$-direction, and its current 
$\rho_{31}$ in $x$-direction are displayed along the $y$-axis.
The spin densities $\rho_{10}$ and $\rho_{20}$ are zero, thus also the
magnetization is correctly aligned along the $z$-axis by the 
magnetic field. Most important, Fig.\ \ref{rho30}  clearly shows that 
the spin-current $\rho_{31}$ differs from zero. This demonstrates that
spin-currents, which can be correctly described by the presented SCDFT,
are far from negligible in real systems.

In summary, we believe that the presented formalism  and method
provide a general and
sound basis for the 
treatment of all sorts of magnetic and (non-colinear) spin effects in a
density-functional  
framework. 
Furthermore, as mentioned above, also spin-orbit interactions can be
included in a natural way in the formalism \cite{rohra06}.

 We thank the John-von-Neumann Institute of Computing for
computational resources.

\end{document}